\documentclass[a4paper]{article}

\usepackage[english]{babel}
\usepackage[utf8x]{inputenc}
\usepackage[T1]{fontenc}
\usepackage{lipsum}
\usepackage{blindtext}
\usepackage{graphicx,amssymb,amsmath,textcomp}
\usepackage[a4paper,top=3cm,bottom=2cm,left=3cm,right=3cm,marginparwidth=1.75cm]{geometry}
\usepackage{tikz}
\usetikzlibrary{matrix}
\usepackage{amsmath}
\usepackage{graphicx}
\usepackage[colorinlistoftodos]{todonotes}
\usepackage[colorlinks=true, allcolors=blue]{hyperref}

\title{\textbf{An Introduction to Geometric Quantization and Witten's Quantum Invariant}}
\author{Kadri \.{I}lker Berktav \footnote{\textit{E-mail:} berktav@metu.edu.tr} \\ \textit{Department of Mathematics, Middle East Technical University}, \\ \textit{06800 Ankara, Turkey}  } 
\date{\vspace{-5ex}}

\usepackage{amsmath,amsthm}

\newtheorem{theorem}{Theorem}[section]

\theoremstyle{proposition}

\theoremstyle{definition}
\newtheorem{definition}{Definition}[section]
\newtheorem{remark}{Remark}[section]

\numberwithin{equation}{section} 

\begin{document}
\maketitle

\begin{abstract}
 This note, in a rather expository manner, serves as a conceptional introduction to the certain underlying mathematical structures encoding  \textit{the geometric quantization formalism} and \textit{the construction of Witten's quantum invariants}, which is in fact organized in the language \textit{topological quantum field theory}.
\end{abstract}
\tableofcontents
\section{Introduction}
A number of remarkable techniques arising from particular gauge theories in physics have long been incarnated into different branches of mathematics. They have been notably employed  
to study low dimensional topology and geometry in a rather sophisticated way, such as Donaldson theory on four-manifolds \cite{Don}, the work of Floer on the topology of 3-manifolds and Yang-Mills instantons that serves as a Morse-theoretic interpretation of Chern-Simons gauge theory (and hence an infinite-dimensional
counterpart of the classical smooth Morse theory \cite{Floer}, \cite{DFK}, \cite{Ruberman}), and Witten's knot invariants \cite{Wit3} arising from a certain \textit{three-dimensional} Chern-Simons theory. Main motivations of this current discussion are as follows: \textit{(i)} to provide a brief introduction to the notion of quantization, \textit{(ii)} to introduce the \textit{geometric quantization formalism} (GQ) and try to understand how the notion of \textit{quantization} boils down to the study of representation theory of \textit{classical observables} in the sense that one can construct \textit{the quantum Hilbert space} $\mathcal{H}$ and a certain Lie algebra homomorphism, and \textit{(iii)} to elaborate in a rather intuitive manner the quantization of Chern-Simons theory together with a brief discussion of a TQFT in the sense of Atiyah \cite{Atiyah} and the language of category theory (cf. \cite{Stacks}, \cite{Vakil}) that manifestly captures the essence of TQFT. With this formalism in hand, we shall investigate Witten's construction of quantum invariants \cite{Wit3} in three-dimensions, and where geometric quantization formalism comes into play.  
\vspace{5pt}

\noindent \textbf{Acknowledgments. }This is \textit{an extended version} of the talk given by the author at \textit{the Workshop on Mathematical Topics in Quantization}, Galatasaray University, Istanbul, Turkey in 2018. The shorter version, on the other hand, will appear in the proceedings of this workshop. This note consists of introductory materials to the notion of geometric quantization based on a series of lectures, namely \textit{Geometric Quantization and Its Applications}, delivered by the author as a weekly seminar/lecture at theoretical physics group meetings organized by Bayram Tekin at the Department of Physics, METU, Spring 2016-2017. Throughout the note, we do not intend to provide neither original nor new results related to subject that are not known to the experts. The references, on the other hand, are not meant to be complete either. But we hope that the material we present herein provides a brief introduction and a na\"{\i}ve guideline to the existing literature for non-experts who may wish to learn the subject. For a quick and accessible treatment to the geometric quantization formalism, including a short introduction to symplectic geometry, see \cite{Blau}, \cite{BDV} or \cite{Honda}. \cite{daS}, on the other hand, provides pedagogically-oriented complete treatment to symplectic geometry. The full story with a more systematic formulation is available in \cite{Woodhouse} and \cite{Brian}.  Finally, I'd like to thank Özgür Kişisel and Bayram Tekin for their comments and corrections on this note and I am also very grateful to them for their enlightening, fruitful and enjoyable conversations during our regular research meetings. Also, I would like to thank organizers and all people who make the event possible and give me such an opportunity to be the part of it.  
\section{Quantization in What Sense and  GQ Formalism}
We would like to elaborate  the notion of geometric quantization in the case of quantization of classical mechanics. Recall that observables in classical mechanics with \textit{a phase space} $(X,\omega)$, a finite dimensional symplectic manifold,  form a \textit{Poisson algebra} with respect to \textit{the Poisson bracket} $\{ \cdot , \cdot \}$ on $C^{\infty}(X)$ given by 
\begin{equation} \label{defn of poisson bracket}
\{f,g \} := -w(X_f,X_g)=X_f (g) \ \ for \ all  \ f,g \in C^{\infty}(X),
\end{equation} where  $X_f$ is \textit{the Hamiltonian vector field associated to} $f$ defined implicitly as \begin{equation} \label{defn of Hamiltonian vector field}
\imath_{ X_f} \omega=df.
\end{equation} Here, $ \imath_{ X_f} \omega $ denotes \textit{the contraction} of a 2-form $\omega$ with the vector field $X_f$ in the sense that 
\begin{equation}
\imath_{ X_f}  \omega \ (\cdot):= \omega(X_f, \cdot).
\end{equation}Employing canonical/geometric quantization formalism (cf. \cite{Honda}, \cite{Blau}, \cite{Woodhouse}, \cite{Brian}), the notion of quantization boils down to the study of representation theory of classical observables in the sense that one can construct the quantum Hilbert space $\mathcal{H}$ and a Lie algebra homomorphism \footnote{\textit{A Lie algebra homomorphism} $\beta:\mathfrak{g} \rightarrow \mathfrak{h}$ is a linear map of vector spaces such that $\beta([X,Y]_{\mathfrak{g}})=[\beta(X),\beta(Y)]_{\mathfrak{h}}$. Keep in mind that, one can easily suppress the constant "-$i\hbar$" in \ref{quantum cond.} into the definition of $\mathcal{Q}$ such that the quantum condition \ref{quantum cond.} becomes the usual compatibility condition that a Lie algebra homomorphism satisfies.} \begin{equation}
\mathcal{Q}:\big(C^{\infty}(X),\{ \cdot, \cdot \}\big)\longrightarrow \big(End(\mathcal{H}),[\cdot , \cdot ]\big)
\end{equation}
together with \textit{the Dirac's quantum condition}: $ \forall $ $f,g\in C^{\infty}(X)
$ we have
\begin{equation} \label{quantum cond.}
[\mathcal{Q}(f), \mathcal{Q}(g)]=-i\hbar \mathcal{Q}\big(\{f ,g \}\big)
\end{equation} where $[\cdot, \cdot]$ denotes the usual commutator on $End(\mathcal{H})$.
\vspace{5pt}

\noindent A primary motivation of this part is to understand how to associate manifestly a suitable Hilbert space $\mathcal{H}$ to a given symplectic manifold $ (M, \omega )$ of dimension $2n$ together with its \textit{Poisson algebra} $ \big(C^{\infty}(M),\{ \cdot , \cdot \}\big) $ in accordance with a certain set of \textit{quantization} axioms given as follows:
\vspace{-18pt}
\begin{quote} 

\begin{definition} \label{defn of quantum system}
(cf. \cite{Blau}, \cite{BDV})
	Let $ (M, \omega )$ be the classical phase space and $\mathcal{A}$ a subalgebra of $ C^{\infty}(M) $. \textit{The quantum system} $ \big(\mathcal{H},\mathcal{Q}\big) $ \textit{associated to} $ \big(M,C^{\infty}(M)\big) $ consists of the following data:
	\begin{enumerate}
		\item A complex separable Hilbert space  $ \mathcal{H} $ where its elements $\psi$ are called the \textit{quantum wave functions} and the rays $ \{\lambda\psi : \lambda \in \mathbb{C}\} $ are \textit{the quantum states}.
		\item For each $f \in \mathcal{A}$, $\mathcal{Q}(f) $ is a self-adjoint $\mathbb{C}$-linear map on $\mathcal{H}$ such that $\mathcal{Q}$ sends the function $f=1$ to the identity operator $id_{\mathcal{H}} \in End(\mathcal{H}).$
		\item \textit{The quantum condition} \ref{quantum cond.} for  $f,g \in \mathcal{A}$.
		\item \textit{The irreducibility condition}: If $\{f_1,...,f_n \}$ is a \textit{complete set} of observables in $\mathcal{A}$, i.e. a function $g\in \mathcal{A}$ commuting with all $f_i$'s must be constant: \begin{equation}
		\{g,f_i\}=0 \ for \ all \ i \Leftrightarrow g=c \ for \ some \ c \in \mathbb{C},
		\end{equation}  then so is the set $\{\mathcal{Q}(f_1),...,\mathcal{Q}(f_n)\}$ of corresponding operators.
	\end{enumerate}
\end{definition} 
\end{quote}    

\noindent\textbf{Geometric quantization} (GQ) is a formalism that encodes the construction of the assignment $ \big(\mathcal{H},\mathcal{Q}\big) $ in a well-established manner (cf. \cite{Blau}, \cite{BDV}, \cite{Woodhouse}, \cite{Brian}). In that respects, it enjoys the following properties: 
\begin{enumerate}
	\item GQ is available for any \textit{finite} dimensional symplectic manifold $ (M, \omega )$.
	\item If $ (M, \omega, \mathcal{G}, \mu )$ is a Hamiltonian $\mathcal{G}$-space with the gauge group $\mathcal{G}$ and  the \textit{moment map} $\mu$ (cf. \cite{daS} ch.22), then GQ \textit{remembers} the symmetries of classical system in the sense that the corresponding quantum states form an \textit{irreducible} representation of $\mathcal{G}$ (this is in fact the representation-theoretic interpretation \cite{Woit} of so-called the irreducibility condition stated above).   
\end{enumerate}
GQ is a two-step process: \textit{(i)} \textit{Pre-quantization}, and \textit{(ii)} \textit{the polarization.}
\textit{The fist step} involves the construction of so-called \textit{a prequantum line bundle} $\mathcal{L}$ on $ (M, \omega )$, the description of a \textit{pre-quantum Hilbert space} $\mathcal{H}_{pre}$ as the space $\Gamma(M,\mathcal{L})$ of smooth square-integrable sections of $\mathcal{L}$ , and a \textit{(pre-) assignment} $\mathcal{Q}_{pre}$  as a certain differential operator acting on such sections of $\mathcal{L}$ (cf. Theorem \ref{existence of prequantum line bundle} and Definition  \ref{defn of GQ assingment}).  Note that even if the first step captures almost all necessary constructions related to the axioms in Definition \ref{defn of quantum system}, it satisfies all but one: \textit{the irreducibility condition}. This is where \textit{the second step} comes into play: In order to circumvent such a pathological assignment, which fails to satisfy the irreducibility condition, we need to restrict the space of smooth functions to be quantized in a certain subalgebra $\mathcal{A}$ that \textit{the irreducibility condition} holds as well. This corresponds to a particular choice of a certain  Lagrangian $n$-subbundle $\mathcal{P}$ of $TM$, called \textit{the polarization}, and hence it leads to define the quantum Hilbert space $\mathcal{H}$ as the space $\Gamma_{\mathcal{P}}(M,\mathcal{L})$ of sections of $\mathcal{L}$ which are \textit{covariantly constant} along $\mathcal{P} \subset TM$ (aka the space of $\mathcal{P}$-polarized sections of $\mathcal{L}$). That is,
\begin{equation}
\Gamma_{\mathcal{P}}(M,\mathcal{L})=\{s\in \Gamma(M,\mathcal{L}) : \nabla_{X} s=0, \ X\in \Gamma(M,\mathcal{P})\subset \Gamma(M,TM)\}.
\end{equation}

\noindent\textit{\textbf{A motivational example.}} This example motivates the notion of polarization in a particular case without providing the formal definition of a polarization (for more detail see \cite{Brian}, \cite{Blau}):  Every K\"{a}hler manifold $ (M, \omega, J) $, where for all $ p\in M$, $J: \ p \mapsto J_p \in End(T_p M)$ is \textit{an integrable almost complex structure compatible with the sypmlectic structure} $\omega$, gives rise to \textit{a holomorphic K\"{a}hler polarization} associated to $(M, \omega)$ by setting $\mathcal{P}:= T^{(0,1)}(M)$, the $(-i)$-eigenspace subbundle of the complexified tangent bundle $TM \otimes \mathbb{C}$. Indeed, since the complex structure $J$ is diagonizable, it defines the \textit{splitting} of the complexified tangent bundle $TM \otimes \mathbb{C}$ as follows: For each $p \in M$, 
\begin{equation}
T_pM \otimes \mathbb{C}=T^{(0,1)}_p(M) \oplus T^{(1,0)}_p(M)
\end{equation} where $ T^{(1,0)}_p(M) = \{v\in T_pM \otimes \mathbb{C} : Jv=iv\} $ and $ T^{(0,1)}_p(M) = \{v\in T_pM \otimes \mathbb{C} : Jv=-iv\} $, which are called \textit{$J$-holomorphic (anti-holomorphic resp.) tangent spaces of $M,$} are both \textit{Lagrangian} subspaces of $ T_pM \otimes \mathbb{C} $ such that $ T^{(0,1)}_p(M) \cap T^{(1,0)}_p(M) =\{0\} $. In local coordinates $(U,z_1, z_2, ... , z_{dim_{\mathbb{C}}M})$ with $z_k=x_k+iy_k$ for $k=1,...,dim_{\mathbb{C}}M$, on the other hand, one has
\begin{equation}
T^{(0,1)}_p(M)=span_{\mathbb{C}} \big\{\partial/\partial \bar{z}_k |_p\big\}_{k=1}^{dim_{\mathbb{C}}M} \ \ \ and \ \ \ \ T^{(1,0)}_p(M)=span_{\mathbb{C}} \big \{\partial/\partial z_k |_p  \big\}_{k=1}^{dim_{\mathbb{C}}M},
\end{equation} where $ \partial/\partial \bar{z}_k=\frac{1}{2}(\partial/\partial x_k+i\partial/\partial y_k) $ and $ \partial/\partial z_k=\frac{1}{2}(\partial/\partial x_k-i\partial/\partial y_k) $. In accordance with the above language, therefore, the space $\Gamma_{\mathcal{P}}(M,\mathcal{L})$ of $\mathcal{P}$-polarized sections of $\mathcal{L}$ is defined as
\begin{equation}
\Gamma_{\mathcal{P}}(M,\mathcal{L})=\{s\in \Gamma(M,\mathcal{L}) : \nabla_{\partial/\partial \bar{z}_k} s=0\}.
\end{equation}
Adopting the usual \textit{summation convention}, we consider, for instance, the case where $M:=\mathbb{C}^n$ with the usual coordinates $\{z_k=x_k+iy_k\}_{k=1}^{n}$ and $\mathcal{L}$ the trivial complex bundle on $M$ together with  the standard K\"{a}hler structure on $M$, described by \textit{the K\"{a}hler potential} $\phi$,

\begin{equation}
\omega=\frac{i}{2} \delta^{jk} dz_j\wedge d\bar{z}_k = \frac{i}{2}\partial\bar{\partial} \phi \ \ \ where \ \ \ \phi = \sum_k |z_k|^2,
\end{equation}
and the usual compatible complex structure: $ J(\partial/\partial x_k)=\partial/\partial y_k $ and $ J(\partial/\partial y_k)=-\partial/\partial x_k $. Since $\mathbb{C}^n$ is a \textit{flat} K\"{a}hler,  we also have $ \nabla_{\partial/\partial \bar{z}_k} =\partial/\partial \bar{z}_k $ and hence the space $ \Gamma_{\mathcal{P}}(M,\mathcal{L}) $ becomes
\begin{equation}
\Gamma_{\mathcal{P}}(M,\mathcal{L})=\{s\in C^{\infty}(\mathbb{C}^n) : \dfrac{\partial s}{\partial \bar{z}_k}=0\}
\end{equation}
which is exactly the space of \textit{holomorphic functions} on $M$. Describing a suitable subalgebra $\mathcal{A}$, on the other hand, is a different story \textit{per se}, and this task is beyond scope of the current discussion.
\vspace{5pt}

 The following section serves as an introductory material and consists of underlying mathematical treatment for the step-\textit{(i)}. Step-\textit{(ii)}, on the other hand, is beyond the scope of this note and will be discussed in detail elsewhere (cf. \cite{Blau}, \cite{Brian} or \cite{Woodhouse}).

\section{The Construction of Prequantization }

We first investigate the quantization of observables in classical mechanics with the phase space $(\mathbb{R}^{2n},q_1,...q_n,p_1,...,p_n)$ and the standard symplectic structure $\omega= \delta^{jk} dq_j\wedge dp_k$ as \textit{a prototype example encoding \textit{the wish list} for the quantum system indicated in Definition \ref{defn of quantum system}.} Recall that given a Hamiltonian function $H\in C^{\infty}(\mathbb{R}^{2n})$, its corresponding Hamiltonian vector field $X_H$ (defined implicitly via \ref{defn of Hamiltonian vector field}) is given locally by 

\begin{equation}
X_H= \delta^{jk} \big(\dfrac{\partial H}{\partial p_j} \dfrac{\partial }{\partial q_k} - \dfrac{\partial H}{\partial q_j} \dfrac{\partial }{\partial p_k} \big).
\end{equation}

\noindent Therefore, for the coordinate functions $ q_j $ and $ p_i $ one has \begin{equation}
X_{q_j}= -\dfrac{\partial }{\partial p_j} \ \ and \ \ X_{p_i}= \dfrac{\partial }{\partial q_i},
\end{equation} and hence the set $\mathcal{S}:= \{ q_1,...q_n,p_1,...,p_n\} $ forms a \textit{complete set} (cf. Definition \ref{defn of quantum system}) due to the following relations:
\begin{equation}
\{q_i,p_j\}=-\delta_{ij} \ \ and \ \ \{q_i,q_j\}=0=\{p_i,p_j\}. 
\end{equation}

\noindent Quantization, on the other hand, gives rise to the similar kind of relations given as
 \begin{equation}
\ [ \mathcal{Q}(p_i),\mathcal{Q}(q_j) ]= -i\hbar\delta_{ij} \ \ and \ \  [ \mathcal{Q}(p_i),\mathcal{Q}(p_j)]=0=[ \mathcal{Q}(q_i),\mathcal{Q}(q_j)]
\end{equation} 
which define so-called \textit{the Heisenberg Lie algebra}. By Schur's lemma, for the complete set $ \mathcal{S}$, the irreducibility condition in Definition \ref{defn of quantum system} boils down to finding the irreducible representations of the Heisenberg algebra where such representations (thanks to the Stone–von Neumann theorem, cf. \cite{Brian} ch. 14, \cite{Blau}) are given by the space $L^2(\mathbb{R}^{n})$ of square-integrable functions on $ \mathbb{R}^{n}$ with the action $ \mathcal{Q} $ of $ C^{\infty}(\mathbb{R}^{2n}) $ on $ L^2(\mathbb{R}^{n}) $ defined as follows: For each $\psi \in L^2(\mathbb{R}^{n}) $ and $\textbf{x}=(x_1,...,x_n) $, we define
\begin{equation}
\mathcal{Q}(q_k)(\psi(\textbf{x})):= x_k\psi(\textbf{x}) \ \ and \ \ \mathcal{Q}(p_k)(\psi(\textbf{x})):= -i\hbar\dfrac{\partial \psi}{\partial x_k}(\textbf{x})
\end{equation}
which  exactly recover so-called \textit{the Schrödinger's picture} of quantum mechanics. 
\vspace{10pt}
\begin{remark}
Note that the underlying mathematical structures of the above example are manifestly discussed in the language of representation theory. The geometric approach, on the other hand, is rather na\"{\i}ve in the sense that the corresponding (pre) quantum line bundle $\mathcal{L}$, which will be elaborated below, is just \textit{the trivial complex bundle} with sections $\psi$ being complex-valued smooth functions and the pre-quantum Hilbert space $\mathcal{H}_{pre} $ being the space of smooth square-integrable sections of $\mathcal{L}$. Furthermore, $ \mathbb{R}^{2n}\backsimeq \mathbb{C}^n $ admits the natural K\"{a}hler structure and the polarization mentioned above. 
\end{remark}

Now, we would like to introduce \textit{an appropriate construction generalizing the above prototype example as follows:} Let $(M,\omega)$ be a symplectic manifold of dimension $2n$. Since $\omega$ is closed 2-form, it defines the de Rham class $ [\omega] \in H^2_{dR}(M) $ and hence it follows from Poincar\'{e} lemma that $ \omega $ is locally exact; that is, there is an open cover $\mathcal{U}= \{U_i\} $ of $M$ such that
\begin{equation}
w=dA_i \  on \ U_i \ \ where \ \ A_i \in \Omega^1(U_i).
\end{equation}
\newpage
\noindent If $ [\frac{\omega}{2\pi\hbar}] \in H^2_{dR}(M;\mathbb{Z})$, then one can construct\textit{ a particular complex line bundle} $\mathcal{L}$ with a certain connection $\nabla$ as follows (cf. \cite{Brian} ch. 22-23 or \cite{Blau}):

\begin{enumerate}
	\item Take the cover $\mathcal{U}$ as a local trivializing cover for $\mathcal{L}$ so that $\mathcal{L}|_{U_i}$ is trivial and $ \omega $ is locally exact on each $U_i$; say $w=dA_i  \ on  \ U_i \ \ where \  A_i \in \Omega^1(U_i).$ We define a connection $\nabla$ on each $U_k$ as 
	\begin{equation}
	\nabla:= d-\frac{i}{\hbar}A_k.
	\end{equation} 
	\item \textit{The gauge transformation} for such cover $\mathcal{U}$ (and hence the transition maps) are defined by making use of Poincar\'{e} lemma on the overlap $U_j \cap U_k$ as follows: Consider two local trivializing sections $s_k: U_k \rightarrow \mathcal{L}$ and  $s_j: U_j \rightarrow \mathcal{L}$ of $\mathcal{L}$. Then, on the overlap $U_j \cap U_k$, we have $ dA_j=\omega=dA_k $; i.e.,
	\begin{equation}
	d(A_k-A_j)=0 \ on \ U_j \cap U_k,
	\end{equation} which implies that $(A_k-A_j)  $ is a closed 1-form on $ U_j \cap U_k $ as well, and hence, from Poincar\'{e} lemma, $ A_k-A_j $ is also locally exact. That is, \begin{equation}
	A_k-A_j= df_{kj} \ for \ some \ f_{kj} \in C^{\infty}(U_j \cap U_k),
	\end{equation} which induces the desired gauge transformations (with the symmetry group $S^1$)
	\begin{equation}
	g: U_j \cap U_k \longrightarrow S^1
	\end{equation}  where $ g(x):= e^{-\frac{i}{\hbar}f_{jk}(x)} $ for all $x\in U_j \cap U_k$ and for all $k,j$ (by which one can define \textit{glueing algorithm} for any two given patches along the overlap).
	\item \textit{The corresponding curvature 2-form} $F_A=dA + A\wedge A$ with this \textit{abelian} gauge group can be expressed  locally as follows: On each $U_k$, one has 
	\begin{equation}
	F_{A_k}=-\frac{i}{\hbar}dA_k=-\frac{i}{\hbar}\omega
	\end{equation}   which leads the following theorem.
\end{enumerate}

\begin{theorem} \label{existence of prequantum line bundle}Let $\omega$ be a closed 2-form on $M$ such that $ [\frac{\omega}{2\pi\hbar}] \in H^2_{dR}(M;\mathbb{Z})$, then there exists a complex line bundle $\mathcal{L}$, called \textit{a prequantum line bundle}, with a connection $\nabla$ as constructed above.
	
\end{theorem}

\noindent Theorem \ref{existence of prequantum line bundle} is at the heart of geometric quantization formalism and it gives rise to the following definition formalized in a rather succinct and na\"{\i}ve way (for a complete treatment see \cite{Brian} ch. 23 or \cite{Blau}):

\begin{definition}  \label{defn of GQ assingment} 
	Let $(M,\omega)$ be a symplectic manifold of dimension $2n$	such that $ [\frac{\omega}{2\pi\hbar}] \in H^2_{dR}(M;\mathbb{Z})$, and $\mathcal{L}$ an associated prequantum line bundle with $\nabla$ as in Theorem \ref{existence of prequantum line bundle}. 
	
	\begin{enumerate}
		\item We set $\mathcal{H}_{pre}:=\Gamma(M,\mathcal{L})$, the space of (equivalence classes of) smooth square-integrable sections (with respect to the Liouville measure on $M$)  of $\mathcal{L}$, with a suitable (hermitian) inner product.
		\item GQ assignment 
		$ \mathcal{Q}_{pre}: \big(C^{\infty}(M),\{ \cdot, \cdot \}\big)\longrightarrow \big(End(\mathcal{H}_{pre}),[\cdot, \cdot ]\big)$  is defined by \begin{equation} \label{Qpre} 
		\mathcal{Q}_{pre}(f):= -i\hbar \nabla_{X_f} - f, 
		\end{equation} which provides the required operator \footnote{If we set $ \mathcal{Q}_{pre}(f):=  \nabla_{X_f}-\frac{i}{\hbar} f $, one would have $ \mathcal{Q}_{pre}\big(\{f ,g \}\big)=[\mathcal{Q}_{pre}(f),\mathcal{Q}_{pre}(g)]$. However, we shall always consider the compatibility condition in the form of \ref{quantum cond.} in oder to capture the physical relevance of the subject and make the interpretation more transparent.} satisfying all axioms except the irreducibility condition in Definition \ref{defn of quantum system}. 
	\end{enumerate}
\end{definition}
\begin{remark}
One can easily verify that the GQ assignment $ \mathcal{Q}$ satisfies \textit{the quantum condition} \ref{quantum cond.} by direct computation together with the definition of $F_A$ as follows:  Recall that for all vector fields $X,Y \in \Gamma(M,TM)$, we have 
\begin{equation} \label{defn of curvature}
F_A (X,Y)=[\nabla_X ,\nabla_Y]-\nabla_{[X,Y]},
\end{equation} 

\noindent and it follows from the construction (cf. Theorem \ref{existence of prequantum line bundle}) that we also have

\begin{equation} \label{defn of curvature as symplectic form}
F_{A} (X,Y) =-\frac{i}{\hbar}\omega (X,Y).
\end{equation}
Let $f,g \in C^{\infty}(M)$ and $s\in \mathcal{H}_{pre}$, then one has 
\begin{align} 
	&(1) \ \ [f,i\hbar\nabla_{X_g}]s = -i\hbar(X_gf)s,\label{observation 1} \\
	&(2) \ \ X_fg=\{f,g\}=-\{g,f\}=-X_gf, \label{observation 2} \\
	&(3) \ \  X_{\{f,g \}} = [X_f,X_g] \ \ \ (from \ Cartan's \ formula \ and \ \ref{defn of Hamiltonian vector field}). \label{Cartans formula}
\end{align}
Therefore, from the definition \ref{Qpre} of $   \mathcal{Q}_{pre}(f) $ and  $ \mathcal{Q}_{pre}(g) $, we obtain 
\begin{align}
[\mathcal{Q}_{pre}(f),\mathcal{Q}_{pre}(g)]s &= [-i\hbar\nabla_{X_f}-f, -i\hbar\nabla_{X_g}-g]s \nonumber \\
&= [-i\hbar\nabla_{X_f},-i\hbar\nabla_{X_g}]s + [-f,-i\hbar\nabla_{X_g}]s + [-i\hbar\nabla_{X_f}, -g]s+[f,g]s \nonumber \\
&= -\hbar^2[\nabla_{X_f},\nabla_{X_g}]s + i\hbar \big(X_fg-X_gf\big)s \ \ \ \ \ \ \ \ \ \ \ \ \  (by \ \ref{observation 1}) \nonumber\\
&= -\hbar^2 \big(F_A (X_f,X_g)+\nabla_{[X_f,Y_g]}\big)s + 2i\hbar \{f,g\}s \ \ \ \ ( by \ \ref{defn of curvature} \ and \ \ref{observation 2}) \nonumber \\
&= i\hbar \omega(X_f,X_g)s-\hbar^2\nabla_{[X_f,Y_g]}s+ 2i\hbar \{f,g\}s \ \ \ \ \ \ \ (by \ \ref{defn of curvature as symplectic form} ) \nonumber\\
&= -\hbar^2\nabla_{[X_f,Y_g]}s + i\hbar \{f,g\}s \ \ \ \ \ \ \ \ (by \ \ref{defn of poisson bracket}) \nonumber \\
& = -\hbar^2\nabla_{X_{\{f,g \}}}s + i\hbar \{f,g\}s \ \ \ \ \ \ \ \ \ (by \ \ref{Cartans formula})  \nonumber \\
&= -i\hbar \big(-i\hbar \nabla_{X_{\{f,g \}}} - \{f,g \}\big)s \ \ \ \ (by \ \ref{Qpre} \nonumber) \\
&= -i\hbar \mathcal{Q}_{pre}\big(\{f ,g \}\big)s
\end{align}
which yields the desired quantum condition \ref{quantum cond.}.
\end{remark}

\section{A Review on Chern-Simons Theory}

 To motivate how the above formalism naturally emerges in the context of a particular quantum field theory and enjoy the richness of this language, we shall study the \textit{quantization} of the $SU(2)$ Chern-Simons gauge theory (\cite{Wit3}) on a closed, orientable 3-manifold $ X $ (we may consider, in particular, an  integral homology 3-sphere for some technical reasons \cite{Ruberman}) as a non-trivial prototype example for a \textit{3-TQFT} formalism in the sense of Atiyah \cite{Atiyah} (for a complete mathematical treatment of the subject, see \cite{Mnev}, \cite{Honda}). 
 \vspace{5pt}

 \noindent Main ingredients of this structure are encoded by the theory of principal $ G $-bundles in the following sense: Let $P\rightarrow X$ be a principal $SU(2)$-bundle on $X$,  $\sigma \in \Gamma(U,P)$ a local trivializing section 
 given schematically as
 \begin{equation}
 	\begin{tikzpicture}
 	\matrix (m) [matrix of math nodes,row sep=3em,column sep=4em,minimum width=2em] {
 		P  & P \\
 		\ & X \\};
 	\path[-stealth]
    (m-1-1)	edge  node [above] {$ \bullet SU(2) $}  (m-1-2)
 	(m-1-2) edge node [right] {$\pi$} (m-2-2)
 	(m-2-2) edge [bend left=40] node [left] {$\sigma$} (m-1-2);
 	\end{tikzpicture}
 	\end{equation}
 
 \noindent Note that when $ G=SU(2) $, $P$ is a trivial principal bundle over $X$, i.e.  $P\cong X \times SU(2)$ compatible with the bundle structure, and hence there exists a globally defined nowhere vanishing section $ \sigma \in \Gamma(X,P)$. Assume $\omega$ is a Lie algebra-valued connection one-form on $P$. Let $ A:= \sigma^* \omega $ be its  representative, i.e. the Lie algebra-valued connection 1-form on $X$, called \textit{the Yang-Mills field}. Then the theory consists of the space of \textit{fields}, which is defined to be the infinite-dimensional space $\mathcal{A}$ of all $ SU(2) $-connections on a principal $ SU(2)$-bundle over $X$, i.e. $ \mathcal{A}:=\Omega^1 (X) \otimes \mathfrak{g} $, and the Chern-Simons action funtional $ CS: \mathcal{A} \longrightarrow S^1$ given by
 \begin{equation}
CS(A):=\frac{k}{4\pi} \displaystyle \int \limits_{X} \mathrm{Tr}(A\wedge \mathrm{d} A +\frac{2}{3}  A \wedge A \wedge A), ~~~~ k\in \mathbb{Z},
\end{equation}
together with the gauge group $\mathcal{G}=Map(X,SU(2))$ acting on the space $\mathcal{A}$ as follows: For all $g\in \mathcal{G}$ and $A \in \mathcal{A}$, we set \begin{equation}
g\bullet A := g^{-1}\cdot A \cdot g + g^{-1}\cdot \mathrm{d} g.
\end{equation}
The corresponding Euler-Lagrange equation in this case turns out to be 
\begin{equation}
F_{A}=0,
\end{equation}where $F_{A}=\mathrm{d} A+A \wedge A$ is the $ \mathfrak{g}$-valued curvature two-form on $X$ associated to $A\in\Omega^1 (X) \otimes \mathfrak{g}.$ Furthermore, under the gauge transformation, the curvature 2-form $F_A$ behaves as follows:
\begin{equation}
F_A \longmapsto g \bullet F_A := g^{-1} \cdot F_A \cdot g \ \ for \ all \ g \in \mathcal{G}. 
\end{equation}

\begin{figure}[h]
	\centering
	
	\includegraphics[width=0.6\linewidth]{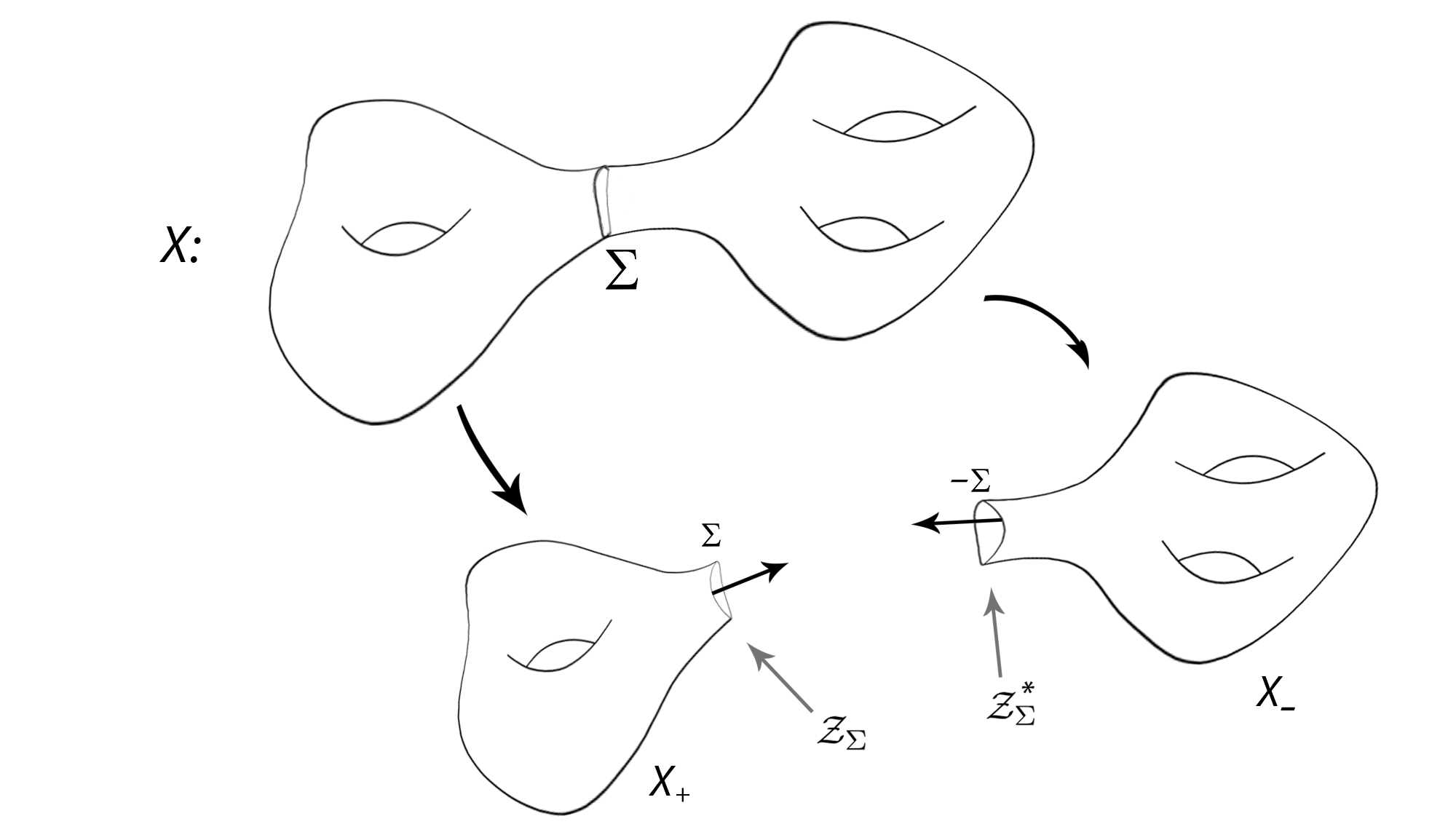}
	\caption{Decomposition of $X$ along a Riemannian surface $\Sigma$.}
	\label{fig:decomposititon}
\end{figure}

Now, in order study the quantization of Chern-Simons theory, we need to adopt the language of \textit{path integral formalism} which will be discussed below (cf. Section \ref{path integral formalism}). The essence of this approach is as follows: In accordance with the axioms of \textit{sigma model} (or those of  \textit{TQFT} as in \cite{Atiyah}, \cite{Honda}, \cite{Mnev}), which will be elaborated succinctly below, we shall consider a decomposition of a closed, orientable 3-manifold $X$ along a Riemannian surface $\Sigma$ (see Figure \ref{fig:decomposititon})  
\begin{equation}
X= (X_{+}\amalg X_{-})/ \Sigma,
\end{equation} 
where $X_{\pm}$ is a compact oriented smooth 3-manifold with boundary $ \partial X_{+} = \Sigma = -\partial X_{-}$ respectively such that $X$ can be obtained by gluing $ X_{+} $ and $ X_{-} $ along their boundaries.

\noindent Then, we would like to study so-called \textit{the partition function $ \mathcal{Z}_X $ assigned to $X$} which essentially captures the probabilistic nature of the quantum Chern-Simons theory and it can be expressed implicitly as a certain pairing (which roughly speaking encodes \textit{the glueing axiom} of QFTs \cite{Honda})
\begin{equation} \label{pairing}
\mathcal{Z}_X= \langle \mathcal{Z}_{X_{+}}, \mathcal{Z}_{X_{-}}\rangle \in \mathbb{C},
\end{equation}
where  $\mathcal{Z}_{\Sigma} $ is \textit{the associated vector space}  together with the natural pairing $\langle-,-\rangle $ on $\mathcal{Z}_{\Sigma} $ such that $ \mathcal{Z}_{X_{+}} \in \mathcal{Z}_{\Sigma}$ and $ \mathcal{Z}_{X_{-}}\in \mathcal{Z}_{\Sigma}^* $. Here $ \mathcal{Z}_{X_{+}} $ and $ \mathcal{Z}_{X_{-}} $ can be considered as "reduced" partition functions associated to each piece $ X_{+} $ and  $ X_{-} $ respectively. Informally speaking, $ \mathcal{Z}_X $ is in fact determined by  data on the boundary via the pairing above with the objects $ \mathcal{Z}_{X_{+}} $ and $ \mathcal{Z}_{X_{-}}. $
\vspace{10pt}

 The following sections will be devoted to \textit{unpackage} the construction of the pairing (\ref{pairing}) and to investigate its relation with low dimensional topology. In order to better understand the underlying mathematical structure encoding the objects like $ \mathcal{Z}_{\Sigma} $ and $\mathcal{Z}_{X_{\pm}}$, we shall briefly discuss the notion of \textit{topological field theory}.

\newpage
\section{TQFT and Category Theory} Before discussing the notion of \textit{topological field theory} in the language of\textit{ category theory}, we first recall how to define a na\"{\i}ve version of TQFT (\cite{Mnev}) in the sense of Atiyah \cite{Atiyah}:

\vspace{-15pt}
\begin{quote} 
\begin{definition}\label{defn of n-tqft}
\textit{A n-TQFT} $\mathcal{Z}$  consists of the following data: \begin{itemize}
		\item For each closed orientable ($n-1$)-manifold $\Sigma$, a vector space $\mathcal{Z}_{\Sigma}$ over $\mathbb{C}$ which is called \textit{the space of states}. Furthermore if $-\Sigma$ denotes reversed-oriented version of $\Sigma$, then one has \begin{equation}
		\mathcal{Z}_{-\Sigma} \cong \mathcal{Z}_{\Sigma}^*
		\end{equation} where $ \mathcal{Z}_{\Sigma}^* $ denotes the linear \textit{dual}  of the vector space $ \mathcal{Z}_{\Sigma} .$
		\item For each compact orientable $n$-manifold $M$ with boundary \begin{equation}
		\partial M= -\Sigma_{in} \amalg \Sigma_{out}
		\end{equation}
		where $M$ is in fact called \textit{$n$-cobordism from $ \Sigma_{in} $ to $ \Sigma_{out} $}, $\mathcal{Z}$ associates a $\mathbb{C}$-linear map of vector spaces
		 \begin{equation}
		\mathcal{Z}_{M}: \mathcal{Z}_{\Sigma_{in}}\longrightarrow \mathcal{Z}_{\Sigma_{out}},
		\end{equation}
		which is called \textit{the partition function}.
		\begin{figure}[h]
			\centering
			\includegraphics[width=0.6\linewidth]{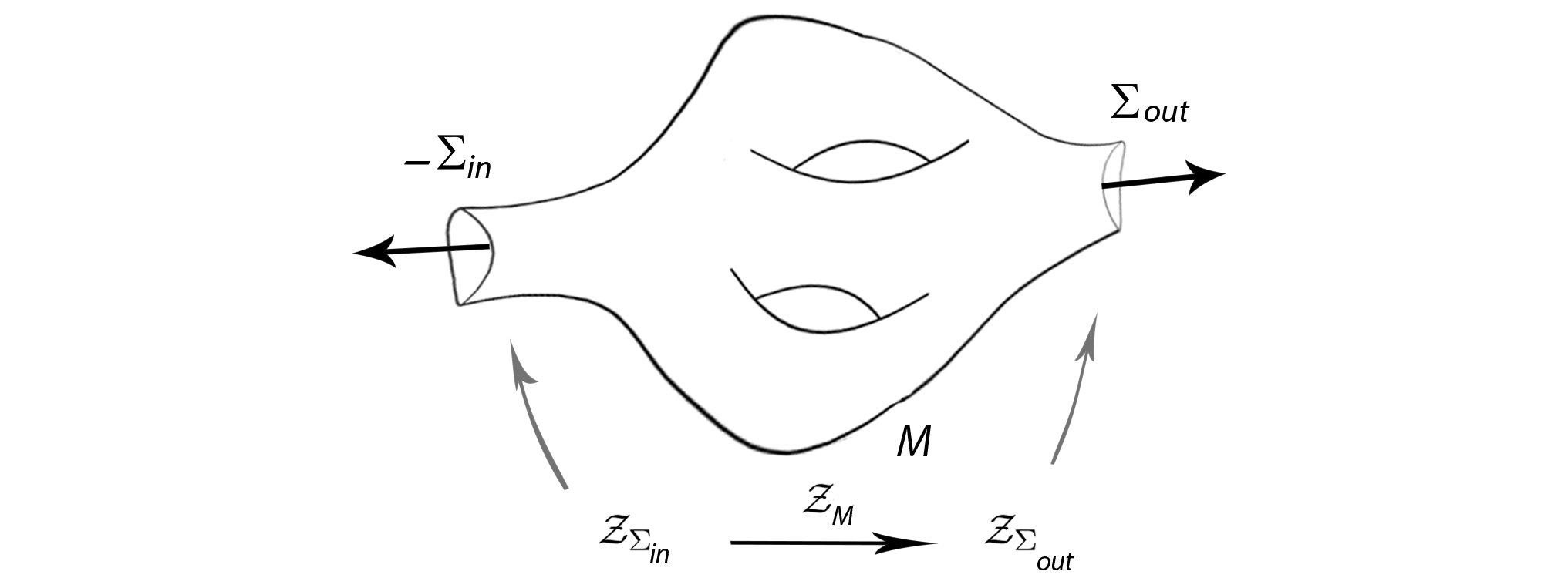}
			\caption{A $n$-cobordism from $ \Sigma_{in} $ to $ \Sigma_{out}. $}
			\label{fig:n-cobordism}
		\end{figure}
		\item If $\Sigma \xrightarrow{f} \Sigma'$ is a diffeomorphism of two closed orientable ($n-1$)-manifolds, then the associated vector spaces are isomorphic: $\mathcal{Z}_{\Sigma} \cong \mathcal{Z}_{\Sigma'}$. If $f$ is orientation-preserving (resp. reversing), then the associated map is $\mathbb{C}$-linear (resp. anti-linear).	
	\end{itemize}
	together with certain \textit{multiplicativity, gluing/composition, normalization} ($\mathcal{Z}_{\emptyset} := \mathbb{C}$) and \textit{compatibility conditions (under diffeomorphisms)} (for a complete definition, see \cite{Mnev}). 	
\end{definition} 
	
\end{quote}

\begin{remark}
The axioms of $ n $-TQFT above in fact encode those of sigma model in a quantum field theory (cf. \cite{Honda}).
With these axioms in hand, note that any closed oriented $n$-manifold $X$ can be realized as a cobordism between $(n-1)$-dimensional empty sets (i.e. a theory from \textit{vacuum} to \textit{vacuum})
 \begin{equation}
\mathcal{Z}_{X}: \mathbb{C}\longrightarrow \mathbb{C}
\end{equation} 
which is defined as a multiplication by some complex number (recall $ \mathcal{Z}_{\emptyset} := \mathbb{C}$).
\end{remark} 
\vspace{5pt}

\noindent\textbf{A digression on main ingredients of category theory.} In this section we would like to introduce a number notions, such as \textit{category, functor between categories} etc., in a rather intuitive manner in the sense that all definitions to be appeared below are given in a relatively na\"{\i}ve sense (for instance, to better articulate the essence of the item without indicating further technical details, we cross our fingers and use repeatedly the phrase \textit{"with certain compatibility conditions"} encompassing certain natural commutative diagrams encoding, for instance, associativity or the behavior under compositions etc...). For a complete mathematical treatment of the subject, see \cite{Vakil} and \cite{Stacks}. 

\newpage
\begin{quote}

\begin{definition} 
	\textit{A category} $\mathcal{C}$  consists of the following data: \begin{itemize}
		\item A \textit{collection} of objects $ Obj (\mathcal{C})$.
		\item For each pair of objects $ A,B \in \mathcal{C} $, there is a \textit{set} $ Mor_{\mathcal{C}}(A,B) $ of morphisms between $ A $ and $ B $. \textit{A morphism} $ f\in Mor_{\mathcal{C}}(A,B)$ is denoted by
		$A\xrightarrow{f}B.$ 
		In particular, for each object $ A \in \mathcal{C} $ there is an identity morphism $ id_{A}: A\rightarrow A $ in $ Mor_{\mathcal{C}}(A,A). $
		\item  For each triple of objects $ A,B,C $ , there is a composition map \begin{equation}
		Mor_{\mathcal{C}}(A,B) \times Mor_{\mathcal{C}}(B,C) \rightarrow Mor_{\mathcal{C}}(A,C),
		\end{equation}
	\end{itemize}
	together with certain \textit{compatibility conditions}. 	
\end{definition}  
\end{quote}
\noindent Some na\"{\i}ve examples of  categories are in order: $ \textit{Top} $ denotes the category of topological spaces with \textit{objects} being topological spaces and \textit{morphisms} being continuous maps between topological spaces. $ \textit{Vect}_{\mathbb{C}} $ denotes the category of vector spaces over $\mathbb{C}$ where objects are vector spaces over $\mathbb{C}$ and morphisms are $\mathbb{C}$-linear maps between such vector spaces.
\vspace{-15pt}
\begin{quote} \begin{definition}
\textit{A (covariant) functor} $\mathcal{F}$: $\mathcal{C} \longrightarrow \mathcal{D}$ between two categories $\mathcal{C} ,\mathcal{D}$ consists of the following data: \begin{itemize}
		\item For objects we have a map $\mathcal{F}: Obj (\mathcal{C})\longrightarrow  Obj (\mathcal{D})$ sending an object $A$ of $\mathcal{C}$ to the object $\mathcal{F}(A)$ of $\mathcal{D}$.
		\item On morphisms, we have a map $ Mor_{\mathcal{C}}(A,B) \longrightarrow Mor_{\mathcal{D}}(\mathcal{F}(A),\mathcal{F}(B))$.
		
	\end{itemize}
	together with certain \textit{compatibility conditions} for compositions and identity morphism, and the existence of the identity functor $ \textit{id}:\mathcal{C} \rightarrow \mathcal{C} $ for each category $\mathcal{C.}$
\end{definition} 

\end{quote}  
With above category-theoretic language in hand, we can re-state the Definition \ref{defn of n-tqft} as follows (cf. \cite{Mnev}):
\vspace{-15pt}
\begin{quote} 
\begin{definition}
A $ n $-TQFT  is a functor $\mathcal{Z}: (Cob_{n},\amalg) \longrightarrow\textit({Vect}_{\mathbb{C}},\otimes) $ of symmetric, monoidal categories where $ (Cob_{n},\amalg) $ denotes the category of \textit{n-cobordisms} with objects being closed orientable ($n-1$)-manifolds and morphisms being $n$-cobordisms.
\end{definition} 

\end{quote}    

\noindent \textit{The end of a digression.}
\vspace{15pt}

 Now, we shall briefly explain where \textit{geometric quantization} comes into play so as to construct the vector space $\mathcal{Z}_{\Sigma}$. To find suitable symplectic manifold to be \textit{quantized}, we need to analyze the critical locus of the Chern-Simons action $CS$ (when restricted to the boundary $\Sigma$). Indeed, we can locally decompose $X$ as $\Sigma \times \mathbb{R}$ where $\Sigma$ is a closed orientable Riemannian surface and $\mathbb{R}$ the time direction. Fixing the gauge condition $A_0=0$, we have the action functional $ CS_{\Sigma}: \mathcal{A}_{\Sigma} \longrightarrow S^1$ of the form
 \begin{equation}
CS_{\Sigma}(A):=\frac{k}{8\pi} \displaystyle \int \mathrm{d}t\int \limits_{\Sigma} \epsilon^{i j} \mathrm{Tr}(A_i \frac{\mathrm{d}}{\mathrm{d}t} A_j), ~~~~ k\in \mathbb{Z},
\end{equation} such that the corresponding field equation is also given by  \begin{equation}
\epsilon^{i j}F^A_{ij}=0,
\end{equation}  which also implies that the connections $A$ on $\Sigma$, which are solutions to the Euler-Lagrange equation, are flat. As stressed in \cite{daS} (ch. 25),  it follows from highly non-trivial theorems of Atiyah and Bott in \cite{AB} that 

\begin{enumerate}
	\item The space $ \big(\mathcal{A}_{\Sigma}, \omega_{\Sigma}, \mathcal{G}\big)  $ is an \textit{infinite-dimensional symplectic manifold} together with a certain choice of $SU(2)$-invariant bilinear form $\langle \cdot, \cdot \rangle $ on its Lie algebra, by which one can define $\omega_{\Sigma}$ manifestly (see \cite{daS} ch.25 for the concrete definition of $ \omega_{\Sigma} $),
	\item Furthermore, the space $ \big(\mathcal{A}_{\Sigma}, \omega_{\Sigma}, \mathcal{G}, \mu\big)  $ is a Hamiltonian $\mathcal{G}$-space with the gauge group $\mathcal{G}$ and  the \textit{moment map} $\mu$ (cf. \cite{daS} ch.25) defined as \textit{the curvature map}, namely \begin{equation}
	\mu: \mathcal{A}_{\Sigma} \longrightarrow LieAlg(\mathcal{G})^*, \ \ A \longmapsto F_A.
	\end{equation}
	\item By using \textit{symplectic reduction theorem} (aka The Marsden-Weinstein-Meyer Theorem, see \cite{daS} ch. 23 for the statement and proof) and results in \cite{AB}, the reduced space  \begin{equation}
	\mathcal{M}_{\Sigma} :=\mu^{-1}(0)/\mathcal{G},
	\end{equation}  which is \textit{the moduli space of flat connections over $\Sigma$ modulo gauge transformation}, turns out to be a compact, finite-dimensional symplectic manifold. Note that the space $ \mathcal{M}_{\Sigma} $ is generically a finite-dimensional symplectic \textit{orbifold} due to the non-freeness of the action of $\mathcal{G}$ on $\mathcal{A}_{\Sigma}$, but in the case where $X$ is a homology 3-sphere and $G=SU(2)$ one can circumvent the pathological quotient by restricting $\mathcal{A}_{\Sigma}$ to a certain dense open subset $\mathcal{A}^* \subset \mathcal{A}_{\Sigma}$ consisting of connections on which $\mathcal{G}$ acts freely (for details see \cite{Ruberman}). 
\end{enumerate}
  With the above observations in hand,  $ \mathcal{M}_{\Sigma} $  serves as a required symplectic manifold to be assigned to $\Sigma$ so that one can construct $\mathcal{Z}_{\Sigma}$ by means of geometric quantization formalism. At the end of the day, therefore,  $\mathcal{Z}_{\Sigma}$ becomes the space of holomorphic sections of a certain complex line bundle (for detailed discussion see \cite{Wit3}). By using the dimensionality of $\mathcal{Z}_{\Sigma}$, on the other hand, one can derive some relations  in terms of \textit{the partition function} $\mathcal{Z}$ (cf. Equation \ref{skeinlike}) such that one can eventually realize that the derived relations turns out to be \textit{the skein relations for the Jones polynomial} in some parameter (see equation \ref{eq:Jonesskeinlike}) if we introduce a knot (oriented) in X. In order to elaborate the last argument, we need to introduce a number of notions that naturally emerge in so-called \textit{the path integral formalism of a quantum field theory} (thought of as a quantum counterpart of the Lagrangian formalism encoding a classical field theory). For an elementary and readable introduction to knot theory, see \cite{Prasolov}.   
\section{The Path Integral Formalism} \label{path integral formalism}  We first recall how to define a na\"{\i}ve and algebro-geometric version of a quantum field theory (\cite{Gw}, \cite{Mnev}) in the path integral formalism: 
\vspace{-15pt}
 \begin{quote} 
 \begin{definition}
\textit{A quantum field theory} on a manifold $X$ consists of the following data: \begin{itemize}
 		\item [\textit{(i)}] the space $ \mathbb{F}_{X}$ of \textit{fields} of the theory defined to be the space $\Gamma(X,\mathcal{F})$ of sections of a particular \textit{sheaf} $ \mathcal{F} $ on $X$,
 		\item [\textit{(ii)}] the action functional $\mathcal{S}: \mathbb{F}_{X}\longrightarrow \mathbb{C}$ \  that captures the behavior of the system under consideration.
 		\item [\textit{(iii)}] An observable $ \varTheta $ defined as a function on $ \mathbb{F}_{X}$:
 		\begin{equation}
 		\varTheta:\mathbb{F}_{X} \longrightarrow \mathbb{C},
 		\end{equation}  
 		\item [\textit{(iv)}] together with its \textit{expectation value} $\langle \varTheta \rangle $ defined by
 		
 		\begin{equation}
 		\langle \varTheta \rangle := \frac{1}{\mathcal{Z}_X} \displaystyle \int\limits_{\phi \in \mathbb{F}_{X}}\varTheta (\phi)e^{iS(\phi)/\hbar} \mathrm{d} \phi,
 		\end{equation}   where $ e^{iS(\phi)/\hbar}  d \phi $ is a \textit{putative} measure on $ \mathbb{F}_{X} $ and \textit{the partition function}  \begin{equation}
 		\mathcal{Z}_X :=\displaystyle \int\limits_{\phi \in \mathbb{F}_{X}}e^{iS(\phi)/\hbar} \mathrm{d} \phi.
 		\end{equation}
 	\end{itemize}
 \end{definition}	
 
 \end{quote}

\vspace{5pt}
  
Now we employ the above formalism for the Chern-Simons theory described at the beginning. We shall study the \textit{quantization} of the $SU(2)$ Chern-Simons gauge theory (\cite{Wit3}) on a closed, orientable 3-manifold $ X $ (in particular, we will take $ X= S^3 $ in a second to make the connection to knot theory more transparent). As before, Let $P\rightarrow X$ be a principal $SU(2)$-bundle on $X$, and $A\in \mathcal{A}:=\Omega^1 (X) \otimes \mathfrak{g} $ the Lie algebra-valued connection 1-form on $X$, then we have
\begin{itemize}
	\item The partition function  \begin{equation}
	\mathcal{Z}_X :=\displaystyle \int\limits_{A \in \mathcal{A}_{X}}e^{iCS(A)/\hbar} \mathrm{d} A
	\end{equation} is a 3-manifold invariant where the integration is a Feynman path integral over all $SU(2)$-connections modulo gauge transformation. Such an invariant can be tractable in accordance with the surgery presentation of given $X$ (see \cite{Wit3}). 
	\item More generally, by introducing a functional $ \varTheta_C (A) $ associated to a connection $A$ on $X$, one can construct an invariant for the \textit{data} $ C $ defining $ \varTheta_C (A) $ as follows 
	\\ \begin{equation}\label{correlation function}
	\mathcal{Z}_{X, \varTheta_C}:= \displaystyle \int\limits_{A \in \mathcal{A}}\varTheta_C (A)e^{iCS(A)/\hbar} \mathrm{d} A
	\end{equation}    
\end{itemize}
The case under consideration to derive knot invariant is that we take $X= S^3$ and $C$, a knot in $X$, together with the structure group $G=SU(2)$ such that  
 \begin{equation}
\varTheta_C (A) := \mathrm{Tr_R} Hol_{A} (C)=\mathrm{Tr}_{R_i} \mathcal{P}e^{i \oint \limits_{C} A}
\end{equation}
where $\mathcal{P}$ denotes \textit{the path ordering} and $Hol_{A} (C) = \{ P_{C} \in GL(P_{x}): P $ is a parallel transport along $C$ defined by $A$\}, the \textit{holonomy group} of $A$ along $C$, and $R$ is a certain irreducible representation of $G$ attached to $C$, which is called a \textit{labeling} of given knot. When we have a \textit{link} $L=\bigcup C_i$, each component $C_i$ is decorated by some irreducible representations $R_i$ of $G$ accordingly and we set \begin{equation}
\varTheta_L(A) :=\prod \varTheta_{C_i} (A)_i, \ \  where \ \ \varTheta_{C_i}(A)_i :=\mathrm{Tr_{R_i}} Hol_{A} (C_i).
\end{equation} Here  $ \varTheta_C (A) $ is called \textit{the Wilson line operator} in the physics literature. In that case, $ \mathcal{Z}_{X, \varTheta_C} $ leads an invariant for $C.$ When we consider a decomposition (Figure \ref{fig:decomposititon}) of $X$ along a Riemannian surface $\Sigma$ (see \cite{Mnev}, \cite{Honda} or \cite{Gw} for details) 
\begin{equation}
X= (X_{+}\amalg X_{-})/ \Sigma,
\end{equation} 
where $X_{\pm}$ is a compact oriented smooth 3-manifold with boundary $ \partial X_{+} = \Sigma = -\partial X_{-}$ respectively such that $X$ can be obtained by gluing $ X_{+} $ and $ X_{-} $ along their boundaries.
Then, in accordance with the axioms of TQFT, we have
\vspace{5pt}

\begin{align}
&\mathcal{Z}_{X_+} :=\displaystyle \int\limits_{A \in \mathcal{A}_{X_+}}e^{iCS(A)/\hbar} \mathrm{d} A \ \in \ \mathcal{Z}_{\Sigma}, \\
&\mathcal{Z}_{X_-} :=\displaystyle \int\limits_{A \in \mathcal{A}_{X_-}}e^{iCS(A)/\hbar} \mathrm{d} A \ \in \ \mathcal{Z}_{-\Sigma}\cong \mathcal{Z}_{\Sigma}^*
\end{align}
\vspace{3pt}

\noindent such that \begin{equation}
\mathcal{Z}_X= \langle \mathcal{Z}_{X_{+}}, \mathcal{Z}_{X_{-}}\rangle \in \mathbb{C},
\end{equation}
where  $\mathcal{Z}_{\Sigma} $ is the vector space associated to $\Sigma$ via \textit{geometric quantization}  together with the natural pairing $\langle-,-\rangle $ on $\mathcal{Z}_{\Sigma} $ such that $ \mathcal{Z}_{X_{+}} \in \mathcal{Z}_{\Sigma}$ and $ \mathcal{Z}_{X_{-}}\in \mathcal{Z}_{\Sigma}^* $.
\vspace{10pt}

Note that the pairing above can be studied more explicitly when we consider \textit{the sigma model}, i.e. a quantum field theory on $X$ with the space of fields being the space $ C_X := Maps(X,N)$ of smooth maps from $X$ to $N$ for some fixed target manifold $ N $, and re-interpreting \textit{the gluing axiom} of sigma model with the help of the usual Fubini's theorem and properties of Feynmann path integrals as follows (see \cite{Honda} for the complete treatment): Let $X, X_{\pm}$ and $\Sigma$ be as above. Then we have

 \begin{align}
  \mathcal{Z}_X &= \displaystyle \int\limits_{\phi \in C_X}e^{iS_X(\phi)/\hbar} \mathrm{d} \phi  \nonumber \\
  &= \displaystyle \int\limits_{\alpha \in C_{\Sigma}} \bigg(\displaystyle \int_{\phi_+ \in C_{X_+}(\alpha)}e^{iS_{X_+}(\phi_+)/\hbar} \mathrm{d} \phi_+ \cdot \displaystyle \int_{\phi_- \in C_{X_-}(\alpha)}e^{iS_{X_-}(\phi_-)/\hbar} \mathrm{d} \phi_-\bigg) \mathrm{d} \alpha \nonumber \\
 &= \displaystyle \int\limits_{\alpha \in C_{\Sigma}}  \mathcal{Z}_{X_{+}} \cdot \mathcal{Z}_{X_{-}}  \mathrm{d} \alpha \nonumber\\
&= \langle \mathcal{Z}_{X_{+}}, \mathcal{Z}_{X_{-}} \rangle, 
\end{align}

\noindent where $C_X (\alpha)$ is a subset of $C_X$ that contains maps $\phi: X \longrightarrow N$ such that $\phi | _{\partial X} = \alpha$, and $\phi_{\pm}$ denote the restriction of $\phi$ to $X_{\pm}$ respectively.
\newpage
\section{The Construction of Witten's Quantum Invariants} 
\begin{figure}[h]
	\centering
	\includegraphics[width=0.65\linewidth]{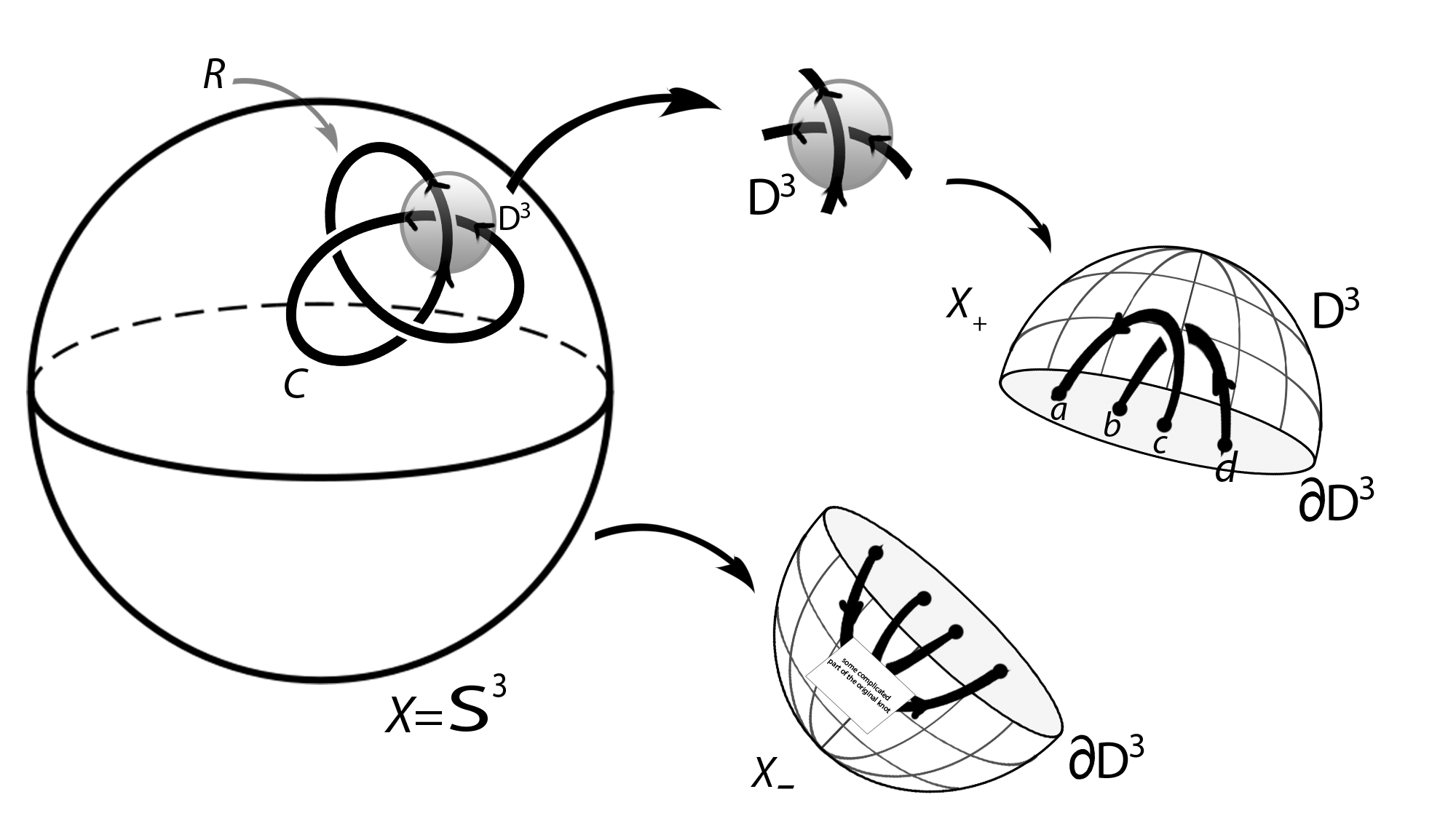}
	\caption{Decomposition of $S^3$ along $\partial D^3=S^2$ containing four marked points $a, b, c, d.$.}
	\label{fig:decomposition with a knot}
\end{figure}

Similar kind of analysis is also applicable when $X$ contains a knot $C$ in a way that after cutting out along a Riemannian surface $\Sigma$, both pieces $X_{\pm}$ involves some part of the knot $C$. Assume $ X $ contains a knot as depicted in Figure \ref{fig:decomposition with a knot}. Consider a ball $D^3$ containing a crossing. Let $\partial D^3=S^2$ play the role of $\Sigma$ as in Figure \ref{fig:decomposititon}, then depending on how pieces of $ X $ (each consists of different part of the original knot) glue back, one can obtain a \textit{non-isotopic} knot, say $L$ (arises from different \textit{brading}, see \cite{Prasolov} for systematic treatment of knot invariants including the computation of certain knot polynomials), and hence different \textit{correlation function}, denoted by $ \mathcal{Z}_{X, \varTheta_L}$ (cf. definition \ref{correlation function}). 
\vspace{5pt}

\begin{remark}
Note that $\partial D^3$ in Figure \ref{fig:decomposition with a knot} is a 2-sphere with some finite number of \textit{marked points} (that is, points with labeling in the sense of above discussion - decorating with certain representations-), and hence we have a vector space $\mathcal{Z}_{\Sigma}$ different from the one that is assigned to 2-sphere without marked points. Furthermore, that sort of vector spaces, the ones that are associated to $\Sigma$ being $ S^2\cong \mathbb{C}P^1 $ with finite number of \textit{marked points} $p_1,...,p_k$, sometimes denoted by $S^2_k$, naturally emerge in other branches of physics and encode some relation between theories in different dimensions, such as the one between \textit{1+1 conformal field theories} (CFTs) and 2+1 dimensional TQFTs. As stressed in \cite{Honda} and \cite{Wit3}, \textit{the space $\mathcal{H}_{S^2_k}$ of conformal blocks} for $S^2_k$ in the context of $ 1+1 $ conformal field theory is the quantum Hilbert space $\mathcal{Z}_{S^2_k}$ obtained by quantizing 2+1 SU(2)-Chern-Simons theory discussed above.  \cite{Schttenloher} provides an elementary introduction to conformal field theory in a particular perspective that is more suited to mathematicians. For an accessible treatment of conformal blocks and the formulation of Witten's knot invariant in the language of conformal field theory, see \cite{Kohno}. Furthermore, \cite{Kohno} and \cite{Schttenloher} also include a systematic treatment for the construction of the space of conformal blocks $\mathcal{H}_{S^2_k}$ for $S^2_k $ and its properties in a way which is essentially based on representation-theoretic approach including so-called \textit{the quantum Clebsch-Gordan condition} and counting the dimension of the space of conformal blocks with the aid of certain combinatorial objects, such as \textit{fusion rules} for surfaces with marked points and \textit{Verlinde formula}. 
\end{remark} 

\noindent With the observations related to the existence of a certain correspondence between 1+1 CFTs and 2+1 TQFTs in hand,  we shall analyze the decomposition depicted in Figure \ref{fig:decomposition with a knot} in detail by adopting the more combinatorial approach appearing in CFT formulation of Witten's knot invariant (see \cite{Kohno}). The sketch of idea is as follows: 
\vspace{10pt}
\begin{itemize}
	\item Without touching anything, i.e. using a diffeomorphims on $\Sigma$ not changing the braiding, such as \textit{the identity map}, if we glue back each pieces $X_{\pm}$ along $\Sigma:=\partial D^3=S^2$ which contains a particular crossing (and hence we in fact have $\Sigma=S^2_4$), then we recover $ X= (X_{+}\amalg X_{-})/ \partial D^3 $ together with the original knot $C$. Otherwise, each configuration differs from each other by a certain diffeomorpmhism of $S^2$ which can be presented in an well-established manner by studying the representation theory of its mapping class group $Map(\Sigma)$. See \cite{Prasolov}, \cite{Kohno} and \cite{Honda} for details.
	\vspace{5pt}
	
	\item Notice that while the piece $X_-$ includes some complicated part of the original knot $C$ (that part is depicted as a \textit{white box} in Figure \ref{fig:decomposition with a knot}), the other one, $X_+$, consists of a part with some \textit{"braiding"} in a sense that each choice of possible braiding corresponds to the one of  "independent" line configurations depicted in Figure \ref{fig:skeinhomfly} that naturally appear in knot theory (see \cite{Prasolov} for more concrete discussion).
	\begin{figure}[h]
		\centering
		\includegraphics[width=0.27\linewidth]{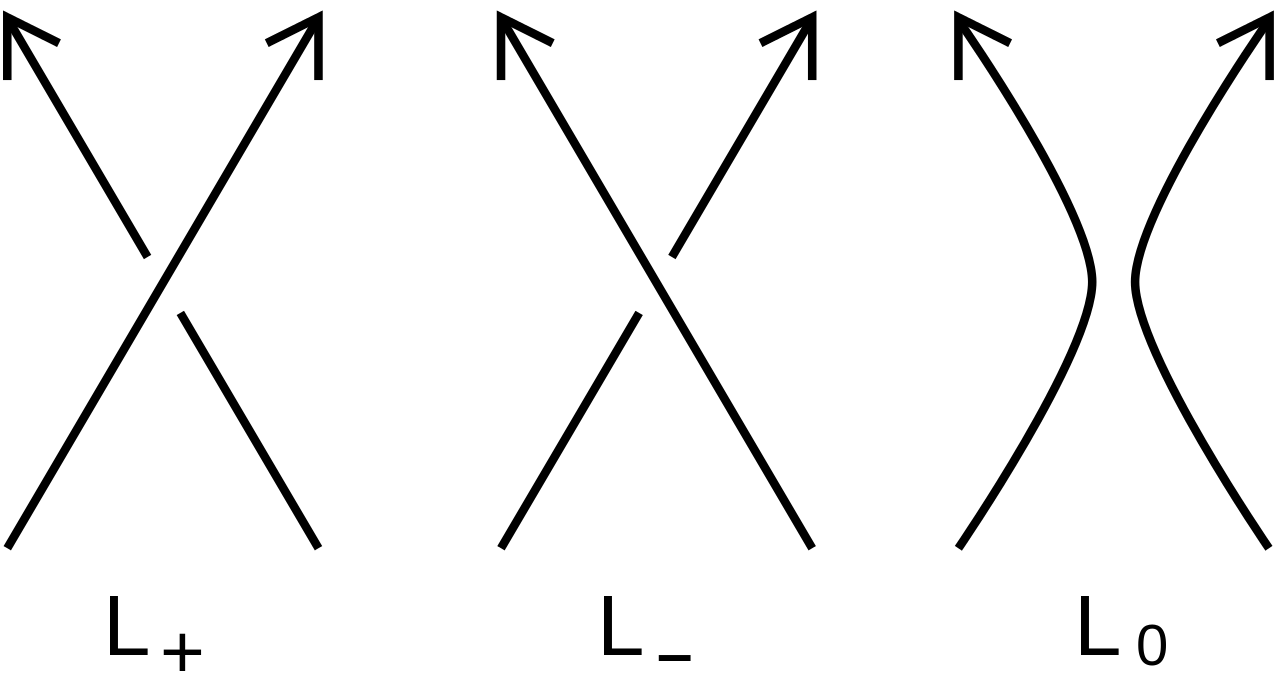}
		\caption{"Independent" line configurations where $ L_{0} $, $ L_{-} $,$ L_{+} $ are the usual notations for zero-crossing, undercrossing and overcrossing resp. in knot theory.}
		\label{fig:skeinhomfly}
	\end{figure}
\item In accordance with the type of line configuration, if we glue $ X_+ $ and $ X_- $ along their boundaries, we can recover either the orinigal knot (including the positive crossing, aka \textit{overcrossing} $ L_+ $, as in Figure \ref{fig:decomposition with a knot}) or the one with \textit{undercrossing} $ L_- $ or the one with \textit{zero-crossing} $ L_0 $. As stressed above, each such line configuration is encoded by a certain diffeomorphism of $ \Sigma $. As a remark, we abuse the notation from now on in the sense that $ L_{0} $, $ L_{-} $ and $ L_{+} $ denote the knots in $X$ obtained from gluing back $ X_{+} $ amd $ X_{-} $ with respect to the choice of line configurations (and hence diffeomorphisms) $ L_{0} $, $ L_{-} $ and $ L_{+} $  respectively.

	\item As outlined in \cite{Kohno}, the choice of braiding of four marked points determines different vectors in the vector space $\mathcal{Z}_{S^2_4}$  associated to the Riemann surface $ S^2_4 $, the 2-sphere with four marked points, in accordance with the axioms of TQFT (or those of sigma model) and the construction provided by GQ formalism. That is, one has the associated vectors
	\begin{equation}
	\mathcal{Z}_{X, \varTheta_{L_-}}, \mathcal{Z}_{X, \varTheta_{L_+}}, \mathcal{Z}_{X, \varTheta_{L_0}} \in \mathcal{Z}_{S^2_4}.
	\end{equation}
	\item Employing representation-theoretic approaches endowed with certain combinatorial techniques such as \textit{fusion rules} and \textit{Verlinde formula} as in \cite{Kohno}, one has the following fact: \begin{equation}
	dim_ {\mathbb{C}} \mathcal{Z}_{S^2_4} \leq 2.
	\end{equation} 

\end{itemize}

\vspace{5pt}

\noindent Due to the finite dimensionality of $\mathcal{Z}_{S^2_4}$, we end up with a certain dependence relation for such vectors corresponding to possible "independent"  configurations, namely \begin{equation} \label{skeinlike}
\alpha \mathcal{Z}_{X, \varTheta_{L_+}}+ \beta \mathcal{Z}_{X, \varTheta_{L_-}}+ \gamma\mathcal{Z}_{X, \varTheta_{L_0}}=0
\end{equation}with some weighted coefficients $\alpha, \beta, \gamma$ which arise from \textit{rational conformal field theory} and manifestly given in \cite{Wit3}. Having computed those coefficients and manipulated the above dependence relation, at the end of the day, we are able to recover \textit{the skein-like relation} defining the Jones polynomial $V(q)$ as follows (\cite{Wit3}):
\begin{equation}\label{eq:Jonesskeinlike}
q^{-1}V(L_+)-qV(L_-)-\big(q^{1/2}-q^{-1/2}\big)V(L_0)=0
\end{equation} where $ q:=e^{2\pi i/(k+2)} $, $k\in \mathbb{Z} $ is the level appearing in the definition of Chern-Simons functional, $V(L_i)$ with $i \in \{0,+,-\}$ denote the Jones polynomial associated to knots with the configurations $ L_{0} $, $ L_{-} $,$ L_{+} $, and we set \begin{equation}
V(q) = \mathcal{Z}_{X, \varTheta_{C}}.
\end{equation}

\begin{remark}
In the physics jargon, evaluating the quantity $ \mathcal{Z}_{X, \varTheta_{C}} $ in fact corresponds to computing \textit{the expectation value of the Wilson line observable associated to the knot $C$ in X}. That essentially gives the 3-dimensional description of knot invariants in terms of 2+1 dimensional $SU(2)$  Chern-Simons theory. Furthermore, if we have a generic closed oriented smooth manifold $ M $, by using the effect of surgery operations (the formal recipe how to obtain $M$ from $S^3$) on the partition function one can effectively evaluate the generalized Jones polynomial for any given knot in $ M.$ This direction is beyond the scope of this section, and for a complete treatment, we again refer to \cite{Wit3}. 
\end{remark} 
\vspace{15pt}

\newpage
\bibliography{xbib}

\end{document}